\documentclass[aps,prl,twocolumn,floatfix,superscriptaddress]{revtex4-1}
\usepackage{amsmath,amssymb,amsfonts}
\usepackage{longtable}
\usepackage{graphicx}
\usepackage{enumitem}
\usepackage{esint}
\usepackage[colorlinks=true,
    linkcolor=blue,
    citecolor=blue,
    urlcolor =blue]{hyperref}

\newcommand{\smean}[1]{\langle{#1}\rangle}

\newcommand{\ket}[1]{\vert{#1}\rangle}

\newcounter{firstbib}

\begin{document}

\title{Nonlinear Quantum Optomechanics via Individual Intrinsic Two-Level Defects}

\author{Tom\'as Ramos}
\affiliation{Institute for Theoretical Physics, University of Innsbruck and\\
Institute for Quantum Optics and Quantum Information of the Austrian Academy of Sciences, 6020 Innsbruck, Austria}
\author{Vivishek Sudhir}
\affiliation{\'{E}cole Polytechnique F\'{e}d\'{e}rale de Lausanne (EPFL), 1015 CH, Lausanne, Switzerland}
\author{Kai Stannigel}
\affiliation{Institute for Theoretical Physics, University of Innsbruck and\\
Institute for Quantum Optics and Quantum Information of the Austrian Academy of Sciences, 6020 Innsbruck, Austria}
\author{Peter Zoller}
\affiliation{Institute for Theoretical Physics, University of Innsbruck and\\
Institute for Quantum Optics and Quantum Information of the Austrian Academy of Sciences, 6020 Innsbruck, Austria}
\author{Tobias J. Kippenberg}
\affiliation{\'{E}cole Polytechnique F\'{e}d\'{e}rale de Lausanne (EPFL), 1015 CH, Lausanne, Switzerland}

\date{\today, Published Version}

\begin{abstract}
We propose to use the intrinsic two-level system (TLS) defect states found naturally in integrated optomechanical devices for exploring cavity QED-like phenomena with localized phonons. The Jaynes-Cummings-type interaction between TLS and mechanics can reach the strong coupling regime for existing nano-optomechanical systems, observable via clear signatures in the optomechanical output spectrum. These 
signatures persist even at finite temperature, and we derive an explicit expression for the temperature at which they vanish. Further, the ability to drive the defect with a microwave field allows for realization of phonon blockade, and the available controls are sufficient to deterministically prepare non-classical states of the mechanical resonator.
\end{abstract}
\maketitle

\emph{Introduction}.-- Cavity optomechanics \cite{KipVah08,MarGir09,MilWoo11} has enabled the preparation of mechanical resonators in states of low phonon occupation
via optomechanical (OM) sideband cooling \cite{Mar07, Wil07, Con10,Pain11,Teu11,Ver12}, and to observe their quantum coherent interaction with light \cite{Ver12}. Further, OM systems have enabled displacement detection at or even below the standard quantum limit \cite{Teu09,Anet10,Wes12}, thereby complementing other mechanics-based sensing applications \cite{Nai09,Gav12}. They have also been proposed for creating macroscopic quantum superpositions \cite{PenBou03} as well as for applications in quantum information \cite{Zhang2003,StanLuk10}.
However, in experiments carried out so far, the interaction between mechanical oscillator and cavity field is effectively linear, while one of the major challenges in the field is to realize nonlinearities at the single phonon level. For example, the intrinsic OM radiation pressure nonlinearity is predicted to enable the generation of non-classical states of light and mechanics \cite{NunGirv11,Rab11}, provided that the single-photon coupling rate exceeds the mechanical frequency and the cavity decay rate. In multi-mode OM systems, the same nonlinearity can be exploited more easily and it has been proposed to use it for enhanced readout \cite{Ludwig2012} and quantum information processing \cite{Stannigel2012}.
\begin{figure}[t]
\includegraphics[width=0.48\textwidth]{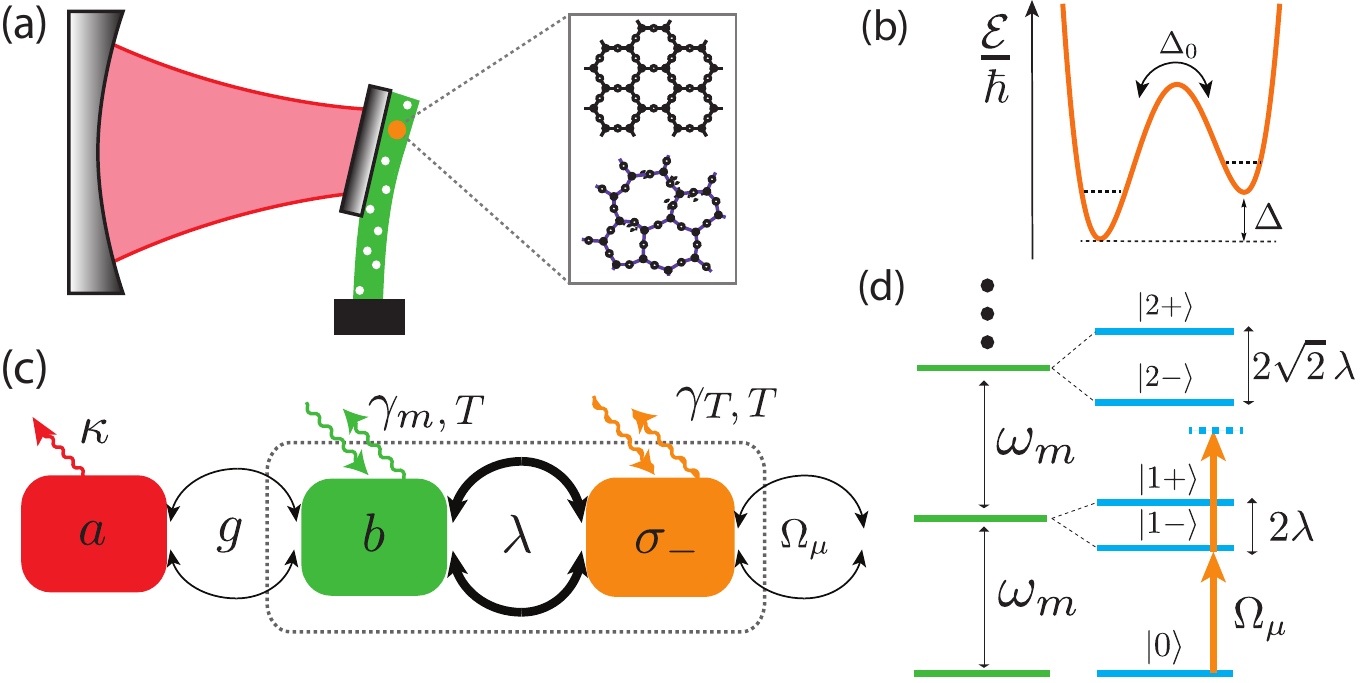}
\caption{(a) Strain coupling of a single two-level system (TLS) defect to an optomechanical system. (b) At low temperature, the defect can be effectively described by two states in a double-well potential, where $\Delta_0$ is the 
tunnel splitting frequency and $\Delta$ the asymmetry frequency. (c) Schematic illustration of decay channels and couplings (see text). Resonator and TLS form a Jaynes-Cummings model (dashed box), exhibiting the characteristic 
anharmonic spectrum shown in (d).\label{fig:schematic}}
\end{figure}

Here we propose an alternative route to render the dynamics of the mechanical oscillator nonlinear at the single quantum level: using its natural coupling to intrinsic structural two-level system (TLS) defects and thereby alleviating the need to functionalize the system [see Fig.\,\ref{fig:schematic}(a)]. Ensembles of TLS defects were first studied in the context of the anomalous and universal low temperature properties of glasses \cite{ZelPol71,AndHalVar72,Phil87,HeuSil93,EnnsHunk}, where they arise from frustration. In experiments involving Josephson junctions, individual TLSs with transition energies distributed well into the GHz 
regime were observed and studied for their role in decoherence \cite{Mart05}. Nevertheless, their comparatively long coherence times, 
and their ability to strongly couple to Josephson junctions via the electric dipole moment have enabled a TLS quantum memory \cite{NeeMart08}. In the same context, the influence of strain on TLSs has been probed recently \cite{Grab12}. However, in the OM setting, TLS ensembles have mainly been studied as a source of decoherence \cite{SeaCas08,RemBle09,Riv11}. In this Letter, we demonstrate theoretically that the coupling of an individual TLS to a localized phonon mode of an OM system can be large enough to exceed the mechanical and TLS dissipation rates, and hence it provides a route to cavity QED-like experiments with single phonons. 
Such experiments have recently been proposed using a different class of defect states,  consisting of donor-acceptor impurity doped silicon \cite{SoyTah11,RusTah12}.
The interaction between TLS and OM system is shown below to be described by a Jaynes-Cummings (JC) Hamiltonian \cite{RemBle09,Tian11}, and induces single-phonon nonlinearities that can be witnessed in the OM cavity output spectrum. Additionally, driving the TLS with microwaves leads to phonon blockade, entailing a mechanical state with sub-Poissonian statistics,
and more complex mechanical states can be engineered using suitable protocols.
Beyond this, our results also apply to other classes of defect states \cite{SoyTah11,RusTah12,Bennett13}, and OM experiments with single defects may mature our yet incomplete understanding of TLSs in glasses \cite{AgaDem12}.

\emph{Strong coupling between resonator and TLS.--} 
We consider OM systems made of silica, where intrinsic TLSs exist due to the amorphous nature of the material, or silicon, where TLS defects reside in the amorphous native (or artificially grown) oxide of the silicon surface \cite{Mor90}.
OM systems can be described by an optical cavity mode $a$ coupled to a co-localized deformational mode $b$ of frequency $\omega_m$, with the Hamiltonian including the intrinsic TLS  given by
\begin{equation}
	H = H_{\rm om} + H_{\rm JC} + H_{{\rm TLS},\mu}.
\end{equation}
Here, $H_{\rm om}=-\hbar\Delta_{L}a^{\dagger}a+ \hbar g(a+a^{\dagger})(b+b^{\dagger})$ describes the standard linearized OM coupling of rate $g$ in a frame rotating at the frequency of the driving laser $\omega_L$, which is detuned from the bare cavity resonance by $\Delta_L$ \cite{KipVah08,MarGir09}. The remaining terms contain the interactions of the mechanical mode with the TLS and the microwave drive of the TLS, as introduced below.

Defects in low-temperature glasses are effectively described by TLSs with tunnel splitting $\hbar\Delta_0$ and asymmetry energy $\hbar\Delta$ \cite{Phil87}, such that the eigenstates are split by $\Delta_T=\sqrt{\Delta^2+\Delta_0^2}$ [see Fig.\,\ref{fig:schematic}(b)]. Since $\Delta$ depends on the strain in the material, the TLS in the OM system couples to the localized phonon mode $b$ \cite{Phil87,RemBle09}. The latter produces a zero-point strain fluctuation on the order of $S_{\rm zpf}=\left(\hbar \omega_m/2 E V_m\right)^{1/2}$, where $E$ is
the Young's modulus of the material and the mechanical mode volume $V_m$ is defined in analogy to optical cavity QED as \cite{supp}
\begin{equation}\label{eq:Vm}
    V_m = \frac{\int T_{ij}(\mathbf{x}) S_{ij}(\mathbf{x})\, d^3 \mathbf{x}}{E S_{kl}(\mathbf{x}_0)S_{kl}(\mathbf{x}_0)}\,.
\end{equation}
Here, $T_{ij}$ and $S_{ij}$ are the tensorial stress and strain profiles, respectively, and repeated indices are summed over. Further, $\mathbf{x}_0$ denotes the point where $S_{ij}S_{ij}$ becomes maximal. Denoting by $\sigma_i$ the Pauli matrices in the TLS eigenbasis, the interaction between resonator and TLS can be approximated by the JC Hamiltonian $H_{\rm JC}=(\hbar\Delta_T/2)\sigma_z + \hbar\omega_m b^\dagger b +
\hbar\lambda (\sigma_{+}b + \sigma_{-}b^\dagger)$, provided that $\lambda\ll\Delta_T\approx\omega_m$ \cite{Tian11}. Here, the TLS-phonon coupling $\lambda$ is given by \cite{supp}
\begin{equation}\label{eq:lambda}
	\lambda \approx \frac{D_T}{\hbar}\frac{\Delta_0}{\Delta_T}
				\left(\frac{\hbar \omega_m}{2E V_m}\right)^{1/2},
\end{equation}
where $D_T$ is the deformation potential \cite{Gold82,AngMan07} and we have neglected factors on the order of one due to the exact position and orientation of the TLS.

In addition to strain, the TLS is also susceptible to classical electromagnetic fields \cite{Phil87,Car94}.
It responds to a coherent microwave drive of Rabi-frequency $\Omega_\mu$ according to $H_{\rm{TLS},\mu}=\hbar\Omega_{\mu}e^{i\omega_{\mu}t}\sigma_{-}+\hbar\Omega_{\mu}^{*}
e^{-i\omega_{\mu}t}\sigma_{+}$.
On the other hand, a static electric field $\mathbf{E}_0$ causes a change in the asymmetry,
$\delta \Delta = 2\mathbf{p}\cdot \mathbf{E}_0/\hbar$, where $\mathbf{p}$ is the dipole moment of the TLS, and thereby changes
the splitting by $\delta \Delta_T = 2(\Delta/\Delta_T)\mathbf{p}\cdot\mathbf{E}_0/\hbar$
and the coupling by a smaller amount $\delta \lambda = -2\lambda (\Delta/\Delta_T^2)\mathbf{p}\cdot\mathbf{E}_0/\hbar$.
Also in the case of artificial donor-acceptor based systems \cite{RusTah12}, similar tuning can be afforded by external electric and magnetic fields.  
\begin{table}[t]
\begin{ruledtabular}
\begin{tabular}{l c c c c}
Mechanical mode & $\omega_m /2\pi$ &  $V_m$ & $\lambda_{\rm max}/2\pi$ & $N_T$ \\
profile         & [GHz]             & [$\mu m^3$] & [MHz] &         \\
\hline \\
\includegraphics[scale=0.065]{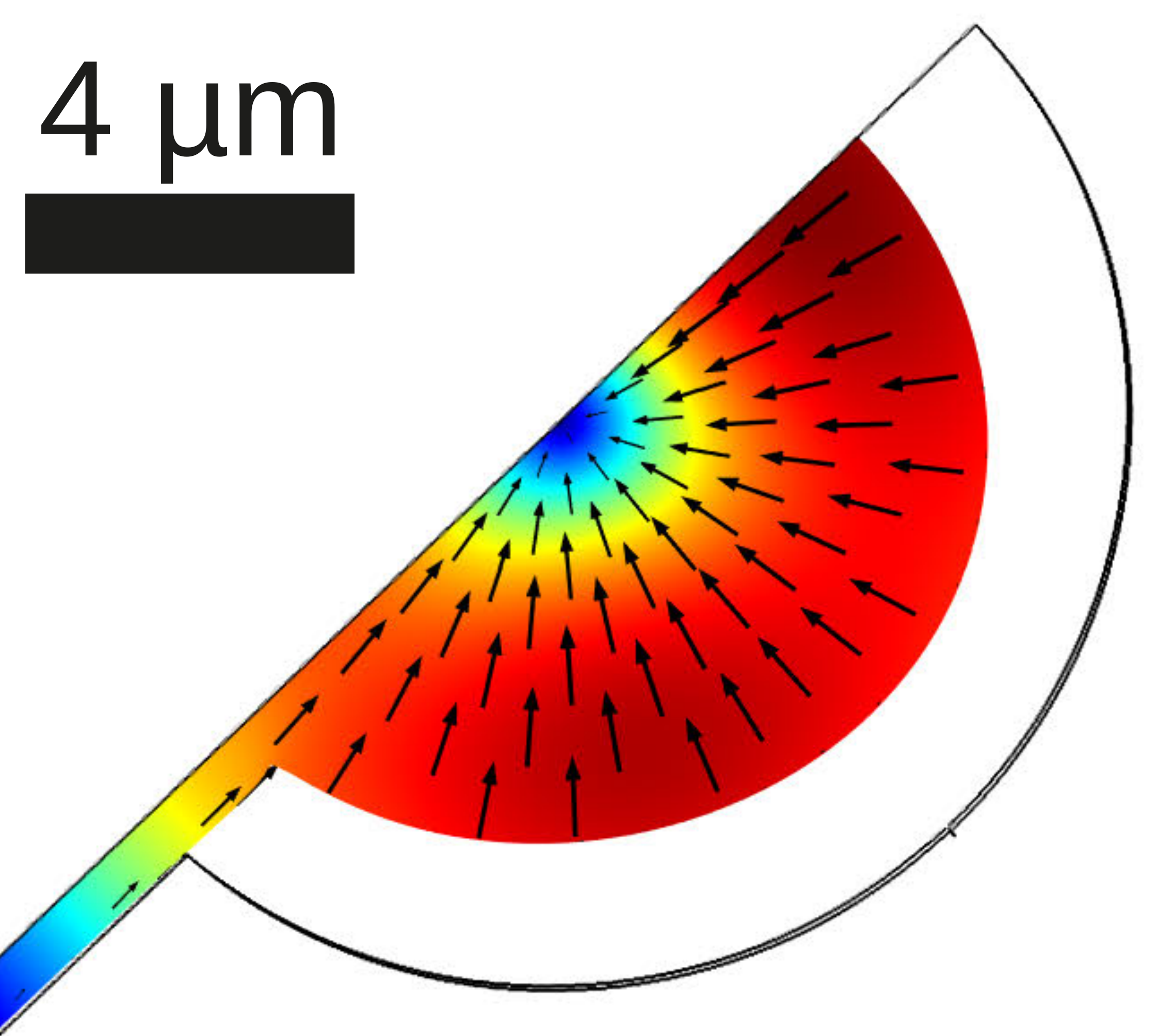} & $0.46$ & $13.46$ & $0.13$ & 0.93 \\
\includegraphics[scale=0.15]{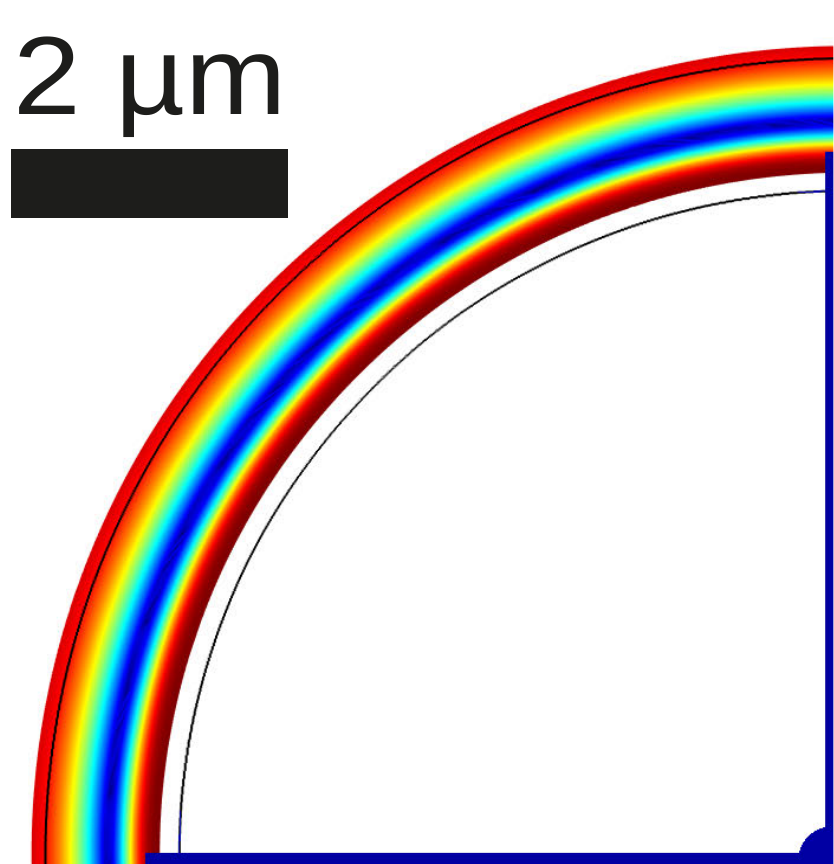} & $1.34$ & $1.32$ & $0.55$ & 0.26 \\
\includegraphics[scale=0.2]{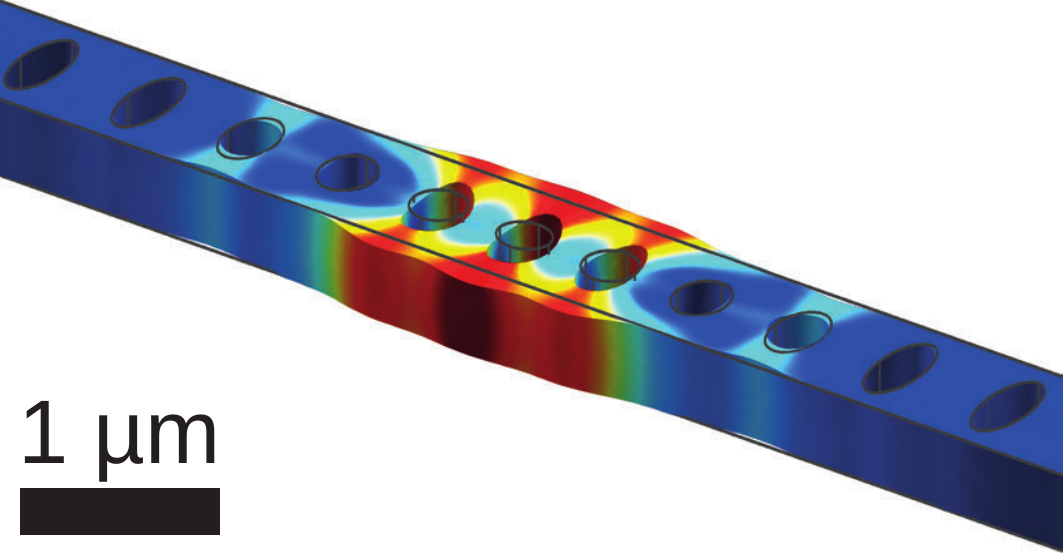} & $5.0$ & $0.01$ & $10.76$ & 0.08 \\
\end{tabular}
\caption{Finite-element simulations of mechanical mode profiles for 
a radial breathing mode of a silica microsphere \cite{VerKimb98}, 
a pinch mode of a Si spoke-anchored microdisc \cite{Tang12}, and 
a localized breathing mode of a Si photonic crystal nanobeam \cite{Pain11} (from top to bottom). Here $\lambda_{\rm max}$ is calculated using Eq.\,\eqref{eq:lambda} with $D_T=1.4\, \mathrm{eV}$ \cite{Gold82}, $\Delta_0 = \Delta_T$, 
and well-known material properties \cite{Zant12}. 
$N_T\approx 0.8 \hbar \lambda_{\rm max} V_T \bar{P}$ denotes the number of relevant TLSs in a volume $V_T$ \cite{supp}, 
i.e. those within a bandwidth $\lambda$ around $\omega_m$ that have $\Delta_0/\Delta_T\gtrsim 0.7$. $\bar P=10^{45}\,\textrm{J}^{-1}\textrm{m}^{-3}$ is the spectral density \cite{RemBle09}. For amorphous silica $V_T$ equals the 
mode volume $V_m$, whereas for silicon, $V_T$ corresponds to the relevant volume of the amorphous native oxide layer \cite{Mor90}.}
\label{tab:results}
\end{ruledtabular}
\end{table}

In Table \ref{tab:results} we display the results of full finite-element simulations of the mechanical modes of three different OM structures. In addition to $\omega_m$, $V_m$ and $\lambda\propto\sqrt{\omega_m/V_m}$, we also show the number $N_T$ of TLSs that couple resonantly and appreciably to the mechanical mode. 
While $N_T\lesssim1$ is desirable in order not to couple to several TLSs, a value of $N_T\ll1$ can be compensated by the above-mentioned tunability, which allows shifting an off-resonant TLS into resonance with the mechanical mode. In particular, already moderate electric fields $|
\mathbf{E}_0|\sim 10^3$\,V/m allow for shifts of $\delta\Delta_T/2\pi\sim 1$ MHz, where we used $|\mathbf{p}|\sim 0.5$ D \cite{Car94} and $\Delta_0/\Delta_T\sim0.9$.
The TLS-phonon couplings $\lambda$ estimated in Table~\ref{tab:results} clearly exceed the typical cryogenic mechanical linewidth $\gamma_m /2\pi \sim 10\ \mathrm{kHz}$ realized in recent experiments \cite{Pain11,Ver12}. On the other hand, typical TLS relaxation rates $\gamma_T/2\pi$ have been measured to be in the range $0.1-5$ MHz for GHz frequencies \cite{Gold82,LisUst10}, and experiments suggest that they dominate over dephasing rates \cite{LisUst10}, which we thus ignore in this work. Using the bulk value $\gamma_T/2\pi\sim1$ MHz at $T\sim 1 \mathrm{K}$ \cite{Phil87}, we obtain $\lambda/\gamma_T\sim 10$ for the case of the nanobeam. The phonon density of states responsible for TLS relaxation is strongly suppressed in the proposed structures as compared to the bulk, which leads us to consider these numbers a worst-case estimate. 

We conclude from the above discussion and the results in Table~\ref{tab:results} that strong coupling to individual defects is feasible for suitably engineered OM systems. While the couplings in the microsphere and microdisc structures are on the verge of the strong coupling regime, the most promising structure is the nanobeam. In the following, we derive the signatures of the TLS-resonator interaction in the OM output spectrum and show that the coupling enables quantum state control of the mechanical resonator in analogy to atomic cavity QED.

\emph{Optomechanical output spectrum}.-- In the absence of the microwave drive ($\Omega_\mu=0$), the system is only driven by thermal and vacuum fluctuations, as described by the master equation
\begin{align}
\label{eq:ME}
\dot\rho &= - \frac{i}{\hbar} [H,\rho] + \mathcal{L}_c\rho + \mathcal{L}_m\rho + \mathcal{L}_T\rho\,,
\end{align}
with Lindblad terms $\mathcal{L}_c\rho = \kappa \mathcal{D}[a]\rho$ for the cavity, and
\begin{align}
\mathcal{L}_T\rho&=\gamma_{T}\left(\bar n_{T}+1\right) \mathcal{D}[\sigma_-]\rho
+ \gamma_{T}\bar n_{T}\mathcal{D}[\sigma_+]\rho,\\
\mathcal{L}_m\rho&= \gamma_m \left(\bar n_{m}+1\right) \mathcal{D}[b]\rho
+ \gamma_m \bar n_{m} \mathcal{D}[b^\dag]\rho,
\end{align}
for TLS and resonator, respectively [see also Fig.\,\ref{fig:schematic}(c)]. Here, $\kappa$ is the cavity energy decay rate, $\bar{n}_{m,T}$ are the Bose occupation numbers corresponding to an environment at temperature $T$ and $\mathcal{D}[x]\rho\equiv x\rho x^\dag- (x^\dag x \rho+ \rho x^\dag x)/2$. We consider the case where the three systems are in resonance ($-\Delta_L\approx\omega_m\approx\Delta_T$), and where the cavity adiabatically follows the resonator dynamics [$\kappa\gg g,\gamma_{T}(\bar{n}_T+1),\gamma_{m}(\bar{n}_m+1)$]. In this regime, the OM coupling leads to cooling of the mechanical resonator and the TLS \cite{Tian11}: For the case $\lambda\ll\kappa$
of relevance in this work, the optically induced mechanical damping rate $A^{(-)}\approx 4g^2/\kappa$ can exceed the corresponding heating rate
$A^{(+)}\approx g^2\kappa/4\omega_m^2$, as well as the rethermalization rate $\gamma_m \bar{n}_m$. In addition, the OM coupling transduces the resonator motion to the cavity output, and thereby enables the observation of the hybridized resonator-TLS subsystem by photodetection. In particular, we consider here the cavity output spectrum $S(\omega)=(\kappa/2\pi)\int_{-\infty}^{\infty}d\tau\ e^{-i(\omega-\omega_L)\tau}\langle a^{\dagger}(\tau)a(0)\rangle_{\rm ss}$, where the angular brackets denote the average in the steady state of Eq. \eqref{eq:ME}.

\begin{figure}[t]
\includegraphics[width=0.50\textwidth]{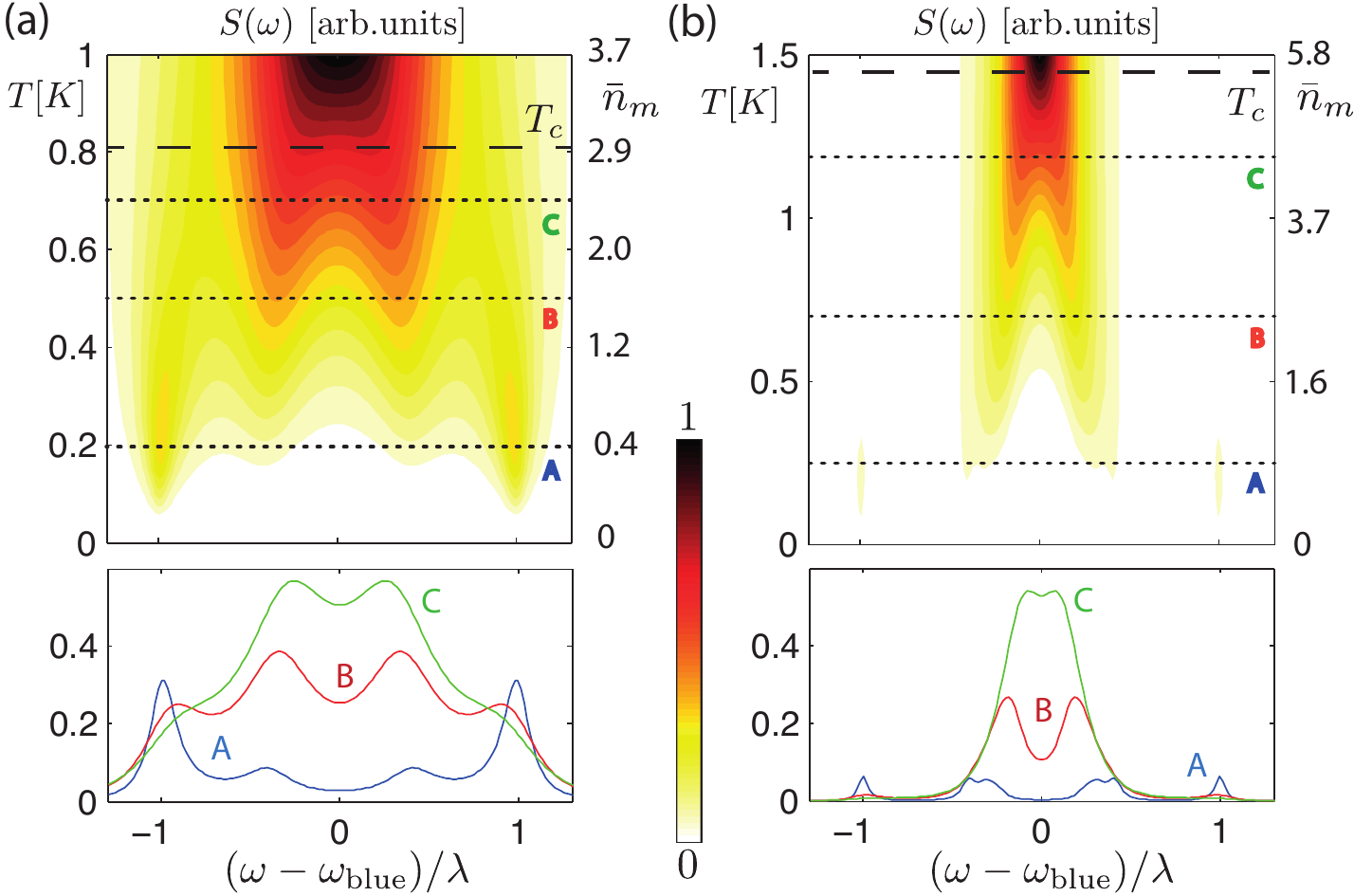}
\caption{Fine structure of the OM output spectrum $S(\omega)$ around the blue sideband at $\omega_{\rm blue}$ as a function of temperature and without a microwave drive field ($\Omega_\mu=0$). The dashed line shows the cross-over temperature according to Eq.\,\eqref{analyticalcross} and the lower panels show cross sections at the temperatures indicated by dotted lines in the top panels. Parameters are $\omega_m/2\pi=5{\rm GHz}$, $\kappa=0.1\omega_m$, $\lambda=2\times 10^{-3}\omega_m$, $\gamma_m=6\times 10^{-6}\omega_m$ and  (a) $g=\lambda$, $\gamma_{T}=\lambda/10$, (b) $g=\lambda/10$, $\gamma_T=\lambda/30$.}
\label{fig:spectra_incoherent}
\end{figure}

In Figure \ref{fig:spectra_incoherent} we display $S(\omega)$ for frequencies around the blue sideband $\omega_{\rm blue}\equiv\omega_L+\omega_m$ as a function of temperature. The spectra show transitions between the levels of the JC Hamiltonian $H_{\rm JC}$ formed by resonator and TLS [dashed box in Fig.\,\ref{fig:schematic}(c)]. In the case of exact resonance ($\omega_m=\Delta_T$), the ground state energy of $H_{\rm JC}$ is zero and the excited states $\ket{n\pm}$ have energies $\omega_{n\pm}=n\omega_m\pm\lambda \sqrt{n}$ ($n=1,2,\ldots$), giving rise to the ``JC ladder'' shown in Fig.\,\ref{fig:schematic}(d). For $T\rightarrow0$ almost all population is in the lowest states of the ladder, such that only the transitions at $\omega=\omega_{\rm blue}\pm\lambda$ between the first rung and the ground state are observed. As temperature increases, higher states get populated, so that transitions between higher rungs located at $\omega=\omega_L+\omega_{n\alpha\beta}$, with
$\omega_{n\alpha\beta}\equiv\omega_{n\alpha}-\omega_{(n-1)\beta}$ ($\alpha,\beta=\pm$), also become visible in the output spectrum. Finally, above a certain cross-over temperature $T_c$ the spectrum consists of a single Lorentzian peak, since the upper and lower branches of the JC ladder resemble two independent highly excited harmonic oscillators \cite{Fink2010}. Note that below $T_c$, temperature actually helps to observe transitions between higher rungs in the JC ladder.

To estimate $T_c$, we adiabatically eliminate the cavity mode and calculate $S(\omega)$ using a secular approximation, which is valid as long as the individual spectral lines do not overlap \cite{TianCarmichael92}. The resulting blue sideband of $S(\omega)$ can be approximated by a sum over Lorentzians $l_\omega(\omega_0,\gamma_0)=\gamma_0^2/(\gamma_0^2+(\omega-\omega_L-\omega_0)^2)$ \cite{supp}, i.e., 
\begin{align}
\label{eq:spectrum}
S_{\rm blue}(\omega)\approx\sum_{n=1}^\infty \sum_{\alpha,\beta=\pm } W^{\rm blue}_{n\alpha\beta}\, l_\omega(\omega_{n\alpha\beta},\gamma_{n\alpha\beta}/2)\,,
\end{align}
where each term corresponds to a transition $\ket{n\alpha}\rightarrow\ket{(n-1)\beta}$  centered at $\omega=\omega_L+\omega_{n\alpha\beta}$. Since $\lambda\ll\kappa$, the widths can be written as $\gamma_{n\alpha\beta}=2(n-1)\bar{\gamma}(\bar{n}+1)+2n\bar{\gamma}\bar{n}+\gamma_T(2\bar{n}_m+1)$, for $n\ge 2$, and $\gamma_{1\alpha+}=\bar{\gamma}(3\bar{n}+1/2)+\gamma_T(2\bar{n}_m+1/2)$. Here, $\bar{\gamma}=\gamma_m+A^{(-)}-A^{(+)}$ is the effective mechanical damping rate and $\bar{n}=(\bar{n}_m\gamma_m+A^{(+)})/\bar{\gamma}$ the corresponding effective mean occupation, as known from standard OM cooling \cite{Wil07,Mar07}. The weights in Eq.\,\eqref{eq:spectrum} are expressed in general as $W^{\rm blue}_{n\alpha\beta}\equiv 2A^{(-)}p_{n\alpha}B^2_{n\alpha\beta}/(\pi\gamma_{n\alpha\beta})$, where $p_{n\alpha}$ is the steady state population of the eigenstate $\ket{n\alpha}$. The JC matrix elements read $B^2_{n\alpha\beta}=[2n-1+2\alpha\beta\sqrt{n(n-1)}]/4$, for $n\geq 2$, and $B_{1\alpha+}^2=1/2$.

As temperature increases and population moves up the JC ladder, transitions between the $\pm$-branches become irrelevant, while the ones within each branch occur closer to $\omega=\omega_{\rm blue}$ and thus contribute to the center of the spectrum. 
We define the cross-over point as the temperature $T_c$ where the separation of the two dominant Lorentzians in Eq.\,\eqref{eq:spectrum} (at each side of $\omega=\omega_{\rm blue}$) equals their width, i.e., where $2\lambda(\sqrt{N}-\sqrt{N-1}) \approx\gamma_{N++}$, with $N$ being the index $n$ for which $W_{n++}^{\rm blue}$ is maximal. If we further restrict the parameters to a regime of experimental interest:
$A^{(+)}\ll\gamma_m \bar{n}_m^c\ll A^{(-)}\lesssim\gamma_T\ll\lambda$
 and $\bar{n}_m^c\gg1$, where $\bar{n}_m^{\rm c}$ is the bath mean occupation at $T_c$, then the cross-over criterion yields \cite{supp}:
\begin{equation}
T_c\approx \frac{\hbar\omega_m}{k_B}\bar n_m^c\approx \frac{\hbar\omega_m}{k_B}\left(\frac{\lambda}{2\gamma_T}\right)^{2/3}\frac{\left(1+2A^{(-)}/\gamma_T\right)\hspace{4mm}}{\left(1+3A^{(-)}/\gamma_T\right)^{2/3}}.\label{analyticalcross}
\end{equation}
Naturally, $T_c$ increases with increasing coupling $\lambda$ and decreases when increasing the TLS linewidth, but additionally Eq.\,\eqref{analyticalcross} shows that $T_c$ can be increased by enhancing the OM cooling rate $A^{(-)}$. The above estimate for $T_c$ agrees well with our numerical simulations, as can be seen from  Fig.\,\ref{fig:spectra_incoherent} (dashed lines). For finite detuning $|\omega_m-\Delta_T|\lesssim \lambda$ of TLS and mechanics the general picture described above remains valid, although the spectra generally become asymmetric.
\begin{figure}[t]
\includegraphics[width=0.5\textwidth]{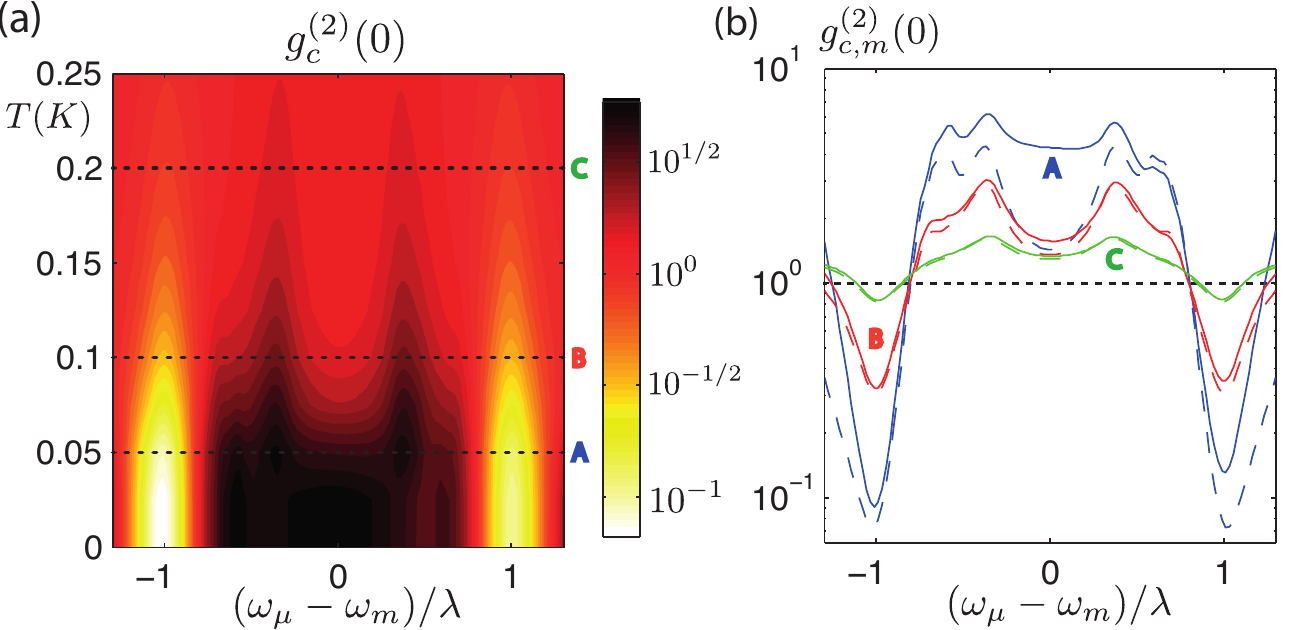}
\caption{Optically probed phonon blockade. (a) $g^{(2)}_c(0)$ of the cavity as a function of drive frequency $\omega_{\mu}$ and temperature $T$ for $|\Omega_\mu|=0.1\lambda$. All other parameters as in Fig.\,\ref{fig:spectra_incoherent}(a). (b) $g^{(2)}_c(0)$ of the cavity (solid) at temperatures indicated by dashed lines in panel (a), and corresponding $g^{(2)}_m(0)$ of the resonator (dashed).}
\label{fig:coherent_g2}
\end{figure}

\emph{Optically probed phonon blockade}.-- By driving the TLS with a weak coherent microwave field ($|\Omega_\mu|\ll\lambda$) we can realize a phonon blockade in analogy to cavity QED \cite{BirKimb05}: if the JC system is cooled to its ground state and the drive is tuned to a transition $\ket{0}\rightarrow\ket{1\pm}$, then the subsequent transition $\ket{1\pm}\rightarrow\ket{2\pm}$ is suppressed provided that $\lambda\gg\gamma_T$, see Fig.\,\ref{fig:schematic}(d). The resulting sub-Poissonian resonator statistics ($g_m^{(2)}(0)\equiv\smean{b^\dag b^\dag b b}/\smean{b^\dag b}^2<1$) persist at finite $T$ if $\exp\left(-\hbar\omega_m/k_B T\right) \ll |\Omega_\mu|^2/\gamma_{1\pm +}^2$, which ensures that the thermal occupation is irrelevant compared to the one due to the drive. Since the cavity operator adiabatically follows the mechanics, i.e. $a(t)\approx -(2ig/\kappa)[b(t)-i(\kappa/4\omega_m)b^{\dag}(t)]+{\rm noise}$ \cite{supp}, we have that $g^{(2)}_m(0)\approx g^{(2)}_c(0)\equiv\smean{a^\dag a^\dag a a}/\smean{a^\dag a}^2$ to zeroth order in $\kappa/\omega_m\ll 1$, such that the phonon blockade can be probed optically. Figure \ref{fig:coherent_g2}(a) displays $g^{(2)}_c(0)$ of the cavity as a function of drive frequency and temperature. For low $T$, one clearly observes the expected antibunching at $\omega_\mu=\omega_m\pm\lambda$, while these features disappear for  higher $T$ due to thermal occupation of the JC ladder. As can be seen from the cuts presented in Fig.\,\ref{fig:coherent_g2}(b), the $g^{(2)}(0)$-function of the cavity is an upper bound for the one of the resonator in the region of pronounced antibunching and therefore, an optical $g^{(2)}_c(0)<1$ indicates phonon blockade.
\begin{figure}[t]
\includegraphics[width=0.47\textwidth]{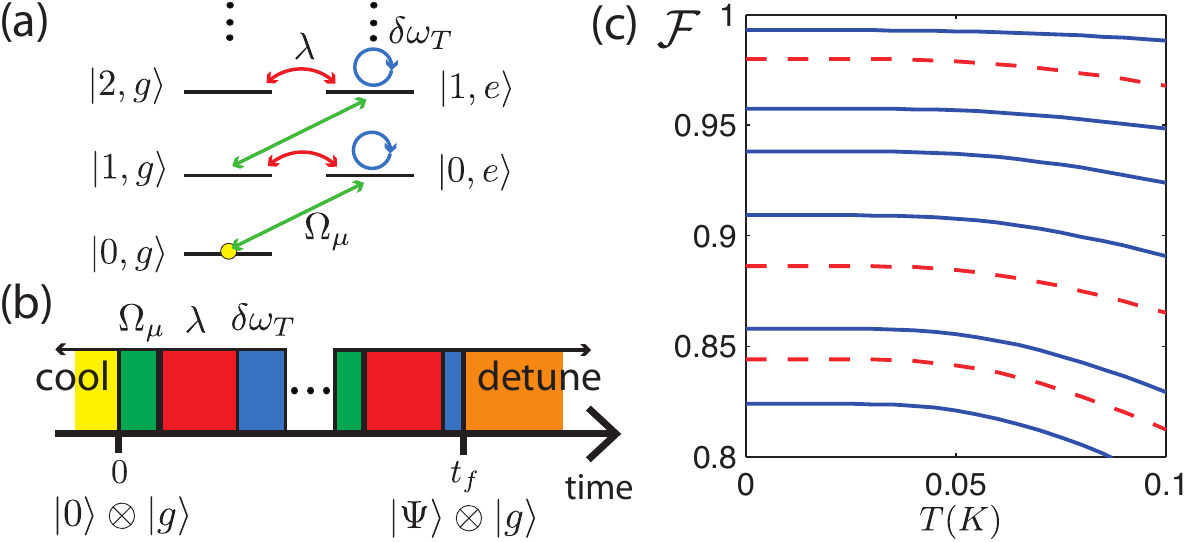}
\caption{Preparation of non-classical resonator states based on Ref. \cite{Law1996}. (a) Level scheme with available tools and (b) general sequence for preparation of a resonator arbitrary state $|\Psi\rangle$ using OM cooling, strong TLS drive ($|\Omega_\mu|\gg\lambda$), free evolution ($\sim \lambda$), strong TLS AC-Stark shift $\delta\omega_T\sim|\Omega_{\mu}|^2/|\omega_{\mu}-\Delta_T|\gg\lambda$, and TLS-resonator detuning. (c) Fidelity ${\cal F}=\langle\Psi_M|\rho(t_f)|\Psi_M\rangle$ as a function of temperature and for increasing $M$ from top to bottom. Dashed curves: $M=1,2,3$ and parameters as in Fig.\,\ref{fig:spectra_incoherent}(a). Solid curves: $M=1,2,3,4,7,9$ and parameters as before, except for $\gamma_T=\lambda/30$. The OM interaction is switched off after the initial cooling.}
\label{fig:state_preparation}
\end{figure}

\emph{Preparation of non-classical states}.-- The possibility to electrically tune and coherently drive the TLS, together with the strong coherent coupling
between TLS and resonator gives rise to the prospect of deterministically preparing quantum states by suitable protocols. As an example, we propose preparing the system close to its ground state by OM cooling and then generating a non-classical resonator state by a scheme analogous to the one of Ref.\,\cite{Law1996}. The necessary sequence is illustrated in Figs.\,\ref{fig:state_preparation}(a,b), and as an example we plot in Fig.\,\ref{fig:state_preparation}(c) the resulting fidelities for preparing the states $|\Psi_M\rangle\equiv\left(|0\rangle+|M\rangle\right)/\sqrt{2}$ ($M=1,2,...$), in the presence of imperfections. Clearly, the fidelity decreases with increasing $M$ and $T$. However, already for $\lambda/\gamma_T=10$ one can achieve $\mathcal{F}>0.95$ for temperatures around $100$mK, which constitutes an exciting avenue for OM systems in existence today.

Work in Innsbruck was supported by the Austrian Science Fund (FWF) through SFB FOQUS and by the EU network AQUTE. T.R. further acknowledges financial support from the BECAS CHILE scholarship program. V.S. and T.J.K. are supported by the ERC starting grant SiMP and the NCCR QSIT program of the Swiss National Science Foundation.

\setcounter{firstbib}{\value{enumiv}}

\cleardoublepage

\section{SI: Coupling between mechanical mode and two-level system}

Starting from the classical description of a solid, we derive the Jaynes-Cummings type interaction of the localized mechanical mode in an optomechanical system with an intrinsic two-level system (TLS) defect.

\subsection{Classical description of a linear and isotropic elastic solid}

The dynamical equation governing the motion of a linear and isotropic elastic solid is given by \cite{SIClelandBook}
\begin{align}
\rho \ddot{u}_i=(\lambda+\mu)\partial_i\partial_j u_j+\mu\partial_j\partial_j u_i,\label{motionequation3d}
\end{align}
where the Einstein summation convetion is used and $u_i=u_i(\mathbf{x},t)$ is the displacement field of the solid with respect to its equilibrium configuration. The quantities $\lambda$ and $\mu$ are the Lam\'e constants, which completely characterize the elastic properties of the linear isotropic solid. They are related to the Young's modulus $E$ and the Poisson's ratio $\nu$ by
\begin{align}
\lambda=\frac{E\nu}{(1+\nu)(1-2\nu)},\qquad
\mu={}&\frac{E}{2(1+\nu)}.
\end{align}
Equation (\ref{motionequation3d}) can be expressed more compactly as
\begin{align}
\ddot{u}_i=\rho^{-1}\alpha_{ijkl}\partial_j\partial_k u_l,\label{motioncompact}
\end{align}
where the elastic moduli tensor $\alpha_{ijkl}$ satisfies the symmetries $\alpha_{ijkl}=\alpha_{jikl}=\alpha_{ijlk}=\alpha_{klij}$ and is given by
\begin{align}
\alpha_{ijkl}=\lambda\delta_{ij}\delta_{kl}
+\mu(\delta_{ik}\delta_{jl}+\delta_{jk}\delta_{il}).\label{explicitmoduli}
\end{align}
Also in terms of $\alpha_{ijkl}$, the generalized Hooke's law for a solid with linear response can be compactly written as
\begin{align}
T_{ij}=\alpha_{ijkl}S_{kl},\label{Hookeslaw}
\end{align}
where $T_{ij}(\mathbf{x},t)$ is the applied stress field on the solid and $S_{ij}(\mathbf{x},t)$ is the correponding strain field, which accounts for the degree of deformation of the solid, defined as
\begin{align}
S_{ij}=\frac{1}{2}\left(\partial_iu_j+\partial_ju_i\right).\label{straindef}
\end{align}

\subsection{Classical Lagrangian for a single mode}

The equation of motion (\ref{motionequation3d}) can be obtained from the Lagrangian
\begin{align}
L=\frac{1}{2}\int d^3x\left(\rho \dot{u}_i\dot{u}_i-\alpha_{ijkl}S_{ij}S_{kl}\right),\label{Lagrangian}
\end{align}
which, owing to the symmetries of $\alpha_{ijkl}$, can be simplified to
\begin{equation}
	L= \frac{1}{2}\int d^3x\left[\rho \dot{u}_i \dot{u}_i - \alpha_{ijkl}(\partial_i u_j)(\partial_k u_l) \right].\label{lagriansimple}
\end{equation}
For the modes of interest, we have free-surface boundary conditions, i.e., the stress must vanish on the surface $\partial V$,
\begin{equation}
	T_{ij}n_j|_{\partial V}=0,\label{boundarycond}
\end{equation}
where $n_i$ is an outwards-pointing unit vector.

Assuming the boundary condition (\ref{boundarycond}), it is shown in Ref.\,\cite{SIAngKuh07} that the differential operator $\rho^{-1}\alpha_{ijkl}\partial_j\partial_k$ on the right hand side of Eq.\,(\ref{motioncompact}) is hermitian and therefore its eigenfunctions form a complete orthonormal set on the domain $V$. In our case, we focus on a single, well-isolated and localized eigenmode of frequency $\omega_m$, which can be described by the ansatz:
\begin{equation}
	u_i (\mathbf{x},t)= \tilde{u}_i (\mathbf{x}) \beta(t),\label{ansatzU}
\end{equation}
such that Eqs.\,(\ref{motionequation3d}) and (\ref{lagriansimple}) reduce to
\begin{align}
0&=\ddot{\beta}+\omega_m^2\beta,\\
0&=\alpha_{ijkl}\partial_{j}\partial_{k}\tilde{u}_l+\rho \omega_m^2 \tilde{u}_i,\label{helmholz}\\
L&=\frac{1}{2}\int d^3x \left[\rho \tilde{u}_i \tilde{u}_i \dot{\beta}^2-\alpha_{ijkl}(\partial_i \tilde{u}_j)(\partial_k \tilde{u}_l)\beta^2\right].\label{ansatzlag}
\end{align}
Completing the total derivative in the second term of Eq.\,(\ref{ansatzlag}), using Eq.\,(\ref{helmholz}), the symmetries of $\alpha_{ijkl}$ and Hooke's law (\ref{Hookeslaw}) for the spatial profile of the stress tensor $\tilde{T}_{ij}(\mathbf{x})=\alpha_{ijkl}\partial_k\tilde{u}_l(\mathbf{x})$, we are left with the Lagrangian,
\begin{equation}
L = \frac{\rho}{2}\left(\dot{\beta}^2 -\omega_m^2 \beta^2 \right)\int d^3x\tilde{u}_i\tilde{u}_i-\frac{\beta^2}{2}\oint_{\partial V} \tilde{u}_i\tilde{T}_{ij}n_j\,dA.
\end{equation}
This form makes explicit that for free-surface and/or fixed-surface boundary conditions, the system is simply described by an harmonic oscillator Lagrangian for $\beta$:
\begin{equation}\label{eq:L_SHO}
	L = \frac{1}{2}m_{\text{eff}}\left(\dot{\beta}^2-\omega_m^2 \beta^2 \right),
\end{equation}
where the effective mass of the mode is defined as \cite{SIPinHeid99,SISch09},
\begin{equation}
	m_{\text{eff}}\equiv\rho \int d^3x \vert \tilde{\mathbf{u}}\vert^2.\label{effectivemassdef}
\end{equation}

\subsection{Quantization of displacement and strain fields}

The quantization of the harmonic oscillator Lagrangian (\ref{eq:L_SHO}) proceeds in the textbook manner and thus the Schr\"odinger picture operator $\hat{\beta}$ can be expressed as
\begin{equation}
	\hat{\beta}= \sqrt{\frac{\hbar}{2m_{\text{eff}}\omega_m}} \left(b+b^\dagger\right),\label{quantizedbeta}
\end{equation}
where $b$ is the destruction operator of the mechanical mode, satisfying $[b,b^\dagger]=1$. The Hamiltonian operator describing the dynamics of $b$ is simply $H_m=\hbar\omega_m\left(b^\dagger b +1/2\right)$.

Based on Eqs.\,(\ref{ansatzU}) and (\ref{quantizedbeta}), the quantized displacement field in the Heisenberg picture reads
\begin{equation}\label{eq:u_operator}
	\hat{u}_i (\mathbf{x},t) = \sqrt{\frac{\hbar}{2m_{\text{eff}}\omega_m}}\, \tilde{u}_i (\mathbf{x}) 
		\left( be^{-i\omega_m t}+b^\dagger e^{i\omega_m t} \right),
\end{equation}
such that the corresponding strain field obtained from Eq.\,(\ref{straindef}) can be expressed as
\begin{equation}
	\hat{S}_{ij}(\mathbf{x},t) = \sqrt{\frac{\hbar}{2m_{\text{eff}}\omega_m}}\, \tilde{S}_{ij} (\mathbf{x}) 
			\left( be^{-i\omega_m t}+b^\dagger e^{i\omega_m t} \right).\label{quantumstrain}
\end{equation}
Here $\tilde{S}_{ij}(\mathbf{x})\equiv(\partial_i \tilde{u}_j+\partial_j \tilde{u}_i)/2$ is the strain profile of the mode of interest. 
We further introduce a normalized strain profile according to $s_{ij}(\mathbf{x})\equiv\tilde{S}_{ij}/S_{\rm norm}$, where $S_{\rm norm} \equiv [\tilde{S}_{ij}(\mathbf{x}_0)\tilde{S}_{ij}(\mathbf{x}_0)]^{1/2}$, and $\mathbf{x}_0$ is the point where $\tilde{S}_{ij}\tilde{S}_{ij}$ becomes maximal. As a result, we have
\begin{equation}
\hat{S}_{ij}(\mathbf{x},t)=S_{\rm zpf}s_{ij}(\mathbf{x})\left( be^{-i\omega_m t}+b^\dagger e^{i\omega_m t} \right),\label{quantumstrainfinal}
\end{equation}
where the zero-point strain fluctuation is given by
\begin{align}
S_{\rm zpf}\equiv\left(\frac{\hbar\omega_m}{2E V_m}\right)^{1/2}.
\end{align}
In analogy to cavity QED \cite{SIHaroche06}, $S_{\rm zpf}$ contains the mode volume $V_m$ of the mechanical mode, which is defined as $V_m\equiv m_{\rm eff}\omega_m^2/ES_{\rm norm}^2$.
Using Eqs.\,(\ref{Hookeslaw}), (\ref{boundarycond}), (\ref{helmholz}), (\ref{effectivemassdef}) and the symmetries of $\alpha_{ijkl}$, we find the relation $m_{\rm eff}\omega_m^2=\int d^3x\ \tilde{T}_{ij}\tilde{S}_{kl}$, such that the mode volume can be written as
\begin{equation}\label{SIeq:Vm}
V_m = \frac{\int \tilde{T}_{ij}(\mathbf{x}) \tilde{S}_{ij}(\mathbf{x})\, d^3 \mathbf{x}}{E \tilde{S}_{ij}(\mathbf{x}_0)\tilde{S}_{ij}(\mathbf{x}_0)}\,.
\end{equation}
This is the result displayed in Eq.\,(2) of the main text.

\subsection{Hamiltonian for the interaction between defects and mechanical mode}\label{TLSmechanicalint}

According to the theory of two-level system (TLS) defects in amorphous solids \cite{SIPhil87,SIAndHalVar72,SIEnnsHunk}, the Hamiltonian describing a single isolated TLS is given by
\begin{equation}
    \bar{H}_{T}= \frac{1}{2}\hbar\Delta\bar{\sigma}_{z}-\frac{1}{2}\hbar\Delta_{0}\bar{\sigma}_{x},\label{TLSHam}
\end{equation}
where $\Delta$ and $\Delta_0$ are the asymmetry frequency and the tunneling amplitude, respectively, and the $\bar{\sigma}_i$ are the Pauli matrices. The TLS couples to electric $\mathbf{E}_0$ and strain $S_{ij}$ fields via its asymmetry, i.e. $\Delta=\Delta(\mathbf{E}_0,S_{ij})$, and for weak fields we write 
\begin{align}
\Delta(\mathbf{E}_0,S_{ij})\approx\Delta+2D_{ij}S_{ij}+2\ \mathbf{p}\cdot\mathbf{E}_0,\label{deltadelta}
\end{align}
where $D_{ij}\equiv(1/2)(\partial\Delta/\partial S_{ij})$ is the deformation potential tensor and $\mathbf{p}\equiv (1/2)(\partial\Delta/\partial \mathbf{E}_0)$ the instrinsic electic dipole moment of the TLS. As discussed in the main text, we exploit the coupling to the electric field for driving the TLS with microwave radiation, but also to tune it by applying a static field.

The deformation potential term in Eq.\,(\ref{deltadelta}) provides the mechanism for coupling the TLS to the localized mechanical resonance of the optomechanical system introduced above. 
To derive the form of this interaction, we replace Eq.\,(\ref{deltadelta}) with $\mathbf{E}_0=0$ into Eq.\,(\ref{TLSHam}) and use the quantum strain field (\ref{quantumstrainfinal}) in the Schr\"odinger picture to obtain
\begin{equation}
    \bar{H}_{T}= \frac{1}{2}\hbar\Delta\bar{\sigma}_z-\frac{1}{2}\hbar\Delta_0\bar{\sigma}_x + \hbar \bar{\lambda}(b+b^\dagger)\bar{\sigma}_z.
\end{equation}
Here the coupling $\bar{\lambda}$ is given by
\begin{equation}
    \bar{\lambda}=\frac{1}{\hbar}S_{\rm zpf}D_{ij}s_{ij}(\mathbf{x}_T),
\end{equation}
where $\mathbf{x}_T$ is the position of the TLS within the optomechanical structure. We are interested in a situation where the eigenfrequency splitting $\Delta_T=\sqrt{\Delta_0^2+\Delta^2}$ of the uncoupled TLS is approximately equal to the resonator frequency in order to obtain a resonant coupling between the two. We further expect the coupling to be much smaller than the free frequencies, such that we concentrate on the situation
\begin{align}
\bar{\lambda}\ll\omega_{\rm m}\approx \Delta_T.
\end{align}
It is thus convenient to change to the eigenbasis (``delocalized picture'') of the uncoupled TLS and to apply the rotating-wave approximation, yielding 
\begin{align}
H_T\approx\frac{1}{2}\hbar\Delta_T\sigma_z
+\hbar\lambda\left(\sigma_+ b+b^{\dagger}\sigma_-\right)\,.\label{hfinal}
\end{align}
Here, the transformed operators read $\sigma_z=(\Delta/\Delta_T)\bar{\sigma}_z-(\Delta_0/\Delta_T)\bar{\sigma}_x$ and $\sigma_x=(\Delta_0/\Delta_T)\bar{\sigma}_z+(\Delta/\Delta_T)\bar{\sigma}_x$, and the general TLS-phonon coupling $\lambda$ is given by
\begin{align}
\lambda\equiv\frac{1}{\hbar}\left(\frac{\Delta_0}{\Delta_T}\right)\left(\frac{\hbar\omega_m}{2E V_m}\right)^{1/2}D_{ij}s_{ij}(\mathbf{x}_T).\label{generalLambdaa}
\end{align}
Finally, neglecting factors of order one due to the exact position and orientation of the TLS, we can approximate Eq.\,(\ref{generalLambdaa}) by
\begin{align}
\lambda\approx\frac{D_T}{\hbar}\left(\frac{\Delta_0}{\Delta_T}\right)\left(\frac{\hbar\omega_m}{2E V_m}\right)^{1/2},\label{simpleLambdaa}
\end{align}
where $D_T$ is the scalar deformation potential that has been measured in the literature (see, e.g., Ref.\,\cite{SIGold82}). Equation (\ref{simpleLambdaa}) is the expression for the TLS-phonon coupling displayed in Eq.\,(3) of the main text.

\subsection{Estimating the number of relevant TLSs}

The number of TLSs with parameters around $\Delta_0$ and $\Delta$ is commonly modeled by the distribution $dN_T=(\hbar V_T\bar{P}/\Delta_0)d\Delta_0d\Delta$, where $V_T$ is the volume of interest and $\bar P$ the spectral density \cite{SIPhil87}. We are interested in the number of TLSs within a bandwidth $\lambda$ around the mechanical frequency $\omega_m$, since these can be considered resonant. Further, the TLS-phonon coupling in Eq.\,(\ref{simpleLambdaa}) should be appreciable, which means that we only count TLSs with $u\equiv\Delta_0/\Delta_T$ above a certain lower cutoff $u_0$. Setting the appropiate integration range and changing to more convenient variables, we thus evaluate
\begin{align}
N_T&=V_T \bar P\hbar
\int_{u_0}^1\textrm{d} u
\int_{\omega_m-\lambda/2}^{\omega_m+\lambda/2}\textrm{d}\Delta_T\,
\frac{1}{u\sqrt{1-u^2}}\\
&=V_T \bar P\hbar\lambda_{\rm max}\int_{u_0}^1\textrm{d} u\frac{1}{\sqrt{1-u^2}}\,,
\end{align}
where we used the fact that $\lambda$ depends on $u$ according to $\lambda=\lambda_{\rm max} u$, with $\lambda_{\rm max}=D_T S_{\rm zpf}/\hbar$. Assuming $u_0=0.7$, the total number of relevant TLSs is estimated to be
\begin{align}
N_T\approx\hbar\lambda_{\rm max} V_T \bar P \times 0.8.
\end{align}
This result is cited in the caption of Table I of the main text; it only moderately depends on the cutoff as long as $u_0\gtrsim 0.5$. For OM systems made of amorphous silica, we identify $V_T$ with the mechanical mode volume $V_m$, since the TLSs reside in the bulk. For silicon systems, we use for $V_T$ the volume of the amorphous native oxide layer  that covers the surface and contains the TLSs \cite{SIMor90}.

\section{SI: Optomechanical output spectrum}

We provide details regarding the calculation of the optomechanical output spectrum in the absence of coherent microwave driving. In particular, we derive the expression for the cross-over temperature $T_c$, at which the TLS signatures in the spectrum vanish. In passing, we present the equations used for analyzing the phonon blockade effect.

The Hamiltonian of the complete system including TLS, resonator, and cavity can be written as $H=H_{\rm om}+H_{\rm JC}$. Here,
\begin{align}
H_{\rm om}=-\hbar\Delta_L a^{\dagger}a+\hbar g(a+a^{\dagger})(b+b^{\dagger})\label{HamOM}
\end{align}
describes the laser-enhanced optomechanical coupling of rate $g$ between the mechanical mode $b$ and the cavity mode $a$ in a frame rotating at the frequency of the driving laser $\omega_L$.  The laser is left implicit throughout this work, which means that $a$ is understood in a corresponding displaced picture \cite{SIWilsonRaeKippenberg2007,SIMar07}. The detuning $\Delta_L$ of the laser from the bare cavity frequency $\omega_c$ reads $\Delta_L\equiv\omega_L-\omega_c$. In addition, the JC Hamiltonian 
\begin{align}
H_{\rm JC}=\hbar\omega_mb^{\dag}b+\frac{1}{2}\hbar\Delta_T\sigma_z+\hbar\lambda(\sigma_{+}b+b^{\dagger}\sigma_{-}),\label{hamiltonianS}
\end{align}
describes the interaction of the mechanical resonator with the TLS.
Apart from coherent interactions we also account for dissipative processes: The cavity is coupled to the electromagnetic vacuum giving rise to an energy decay rate $\kappa$, whereas the mechanical oscillator and the TLS interact with indepedent heat baths at temperature $T$. The corresponding energy decay rates are denoted by $\gamma_m$ and $\gamma_T$, respectively. Again using $\lambda\ll\omega_m\approx\Delta_T$, the dynamics of the total open system is described by the master equation
\begin{align}
\dot{\rho}(t)={}&-\frac{i}{\hbar}[H,\rho]+\kappa{\cal D}[a]\rho\nonumber\\
&+\gamma_m(\bar{n}_m+1){\cal D}[b]\rho+\gamma_m\bar{n}_m{\cal D}[b^{\dagger}]\rho\nonumber\\
&+\gamma_T(\bar{n}_T+1){\cal D}[\sigma_{-}]\rho+\gamma_T\bar{n}_T{\cal D}[\sigma_{+}]\rho,\label{masterequation}
\end{align}
where $\mathcal{D}[x]\rho\equiv x\rho x^\dag- (x^\dag x \rho- \rho x^\dag x)/2$ denotes a Lindbland term, and $\bar n_{m,T}$ are the Bose occupation numbers of the baths, evaluated at the mechanical and TLS frequencies, respectively.

\subsection{Jaynes-Cummings eigenbasis}

The JC Hamiltonian $H_{\rm JC}$ can be exactly diagonalized \cite{SIJCmodel63}. Denoting by $|n,g\rangle$ and $|n,e\rangle$ the states with $n$ phonons and the TLS in its ground or excited state, respectively, the eigenstates $|n\pm\rangle$ of $H_{\rm JC}$ can be written as
\begin{align}
|n\alpha\rangle=C_{n\alpha}|n,g\rangle+S_{n\alpha}|n-1,e\rangle,\ \ \ n\ge 1,\ \alpha=\pm,
\end{align}
and $|0+\rangle=|0,g\rangle$ for the ground state ($|0-\rangle=0$). The coefficients $C_{n\alpha}$ and $S_{n\alpha}$ are explicitly given by $C_{n\alpha}=\alpha\left[(\Omega_n+\alpha\delta\omega)/2\Omega_n\right]^{1/2}$ and $S_{n\alpha}=\left[(\Omega_n-\alpha\delta\omega)/2\Omega_n\right]^{1/2}$, where $\Omega_n=(\delta\omega^2+4n\lambda^2)^{1/2}$ and the TLS-phonon detuning reads $\delta\omega=\omega_m-\Delta_T$. The JC Hamiltonian in its eigenbasis reads
\begin{align}
H_{\rm JC}={}&\sum_{n=0}^{\infty}\sum_{\alpha=\pm}\omega_{n\alpha}|n\alpha\rangle\langle n\alpha|,\label{HJCdiag}
\end{align}
where the eigenfrequencies can be calculated to be
\begin{align}
\omega_{n\alpha}={}&n\omega_m+\frac{1}{2}(\alpha\Omega_n-\delta\omega),\ \ \ n\ge 1,\ \ \alpha=\pm,
\end{align}
and $\omega_{0\pm}=0$. Also in the JC eigenbasis, the operators $b$ and $\sigma_{-}$ can be decomposed as
\begin{align}
b={}&\sum_{n=1}^{\infty}\sum_{\alpha,\beta=\pm}B_{n\alpha\beta}{\cal O}_{n\alpha\beta},\label{descom1}\\
\sigma_{-}={}&\sum_{n=1}^{\infty}\sum_{\alpha,\beta=\pm}\sigma_{n\alpha\beta}{\cal O}_{n\alpha\beta},\label{descom2}
\end{align} 
where the transition operators between the eigenstates of adjacent rungs in the JC ladder are given by
\begin{align}
{\cal O}_{n\alpha\beta}=|(n-1)\beta\rangle\langle n\alpha|.
\end{align}
The matrix elements $\sigma_{n\alpha\beta}$ and $B_{n\alpha\beta}$ read
\begin{align}
B_{n\alpha\beta}={}&C_{(n-1)\beta}C_{n\alpha}\sqrt{n}
+S_{(n-1)\beta}S_{n\alpha}\sqrt{n-1},\label{matrixBs}\\
\sigma_{n\alpha\beta}={}&C_{(n-1)\beta}S_{n\alpha},
\end{align}
for $n\ge 2$. For $n=1$, one has $B_{1\alpha +}=C_{1\alpha}$, $\sigma_{1\alpha +}=S_{1\alpha}$ and the remaining coefficients vanish.  

\subsection{Effective master equation for mechanical mode and two-level system}

As discussed in the main text, we are interested in the situation where the optical cavity adiabatically follows the dynamics of resonator and TLS, giving rise to an optomechanical cooling effect. Based on the conditions
\begin{align}
\label{eq:timescales}
\kappa\gg g,\gamma_{T}(\bar{n}_T+1),\gamma_{m}(\bar{n}_m+1)\,,
\end{align}
we thus derive an effective master equation for the reduced density matrix $\rho_s$ of resonator and TLS by adiabatically eliminating the cavity mode. To this end, we transform the master equation (\ref{masterequation}) to an interaction picture defined by the unitary $U(t)=e^{-i(H_{\rm JC}-\Delta_L a^{\dagger}a)t}$ and express the result as
\begin{align}
\dot{\tilde{\rho}}={}&\tilde{{\cal L}}_c\tilde{\rho}+\tilde{{\cal L}}_{\rm int}\tilde{\rho}+\tilde{{\cal L}}_{\rm JC}\tilde{\rho},\label{masterequation3}
\end{align}
where $\tilde\rho=U^\dag\rho U$ and $\tilde{{\cal L}}_c\tilde{\rho}=\kappa{\cal D}[a]\tilde{\rho}$. In the JC eigenbasis the optomechanical interaction reads
\begin{align}
\tilde{{\cal L}}_{\rm int}\tilde{\rho}={}&-ig\sum_{n\alpha\beta}B_{n\alpha\beta}[(\tilde{a}+\tilde{a}^{\dagger})(\tilde{{\cal O}}_{n\alpha\beta}+\tilde{{\cal O}}_{n\alpha\beta}^{\dagger}),\tilde{\rho}],\label{liouvint3}
\end{align}
where the operators now carry a time-dependence according to $\tilde{a}(t)=e^{i\Delta_L t}a$ and $\tilde{{\cal O}}_{n\alpha\beta}(t)=e^{-i\omega_{n\alpha\beta}t}{\cal O}_{n\alpha\beta}$, with $\omega_{n\alpha\beta}\equiv\omega_{n\alpha}-\omega_{(n-1)\beta}$ being the difference between JC eigenfrequencies of adjacent rungs. The dissipative dynamics of resonator and TLS is described by
\begin{align}
\tilde{{\cal L}}_{\rm JC}\tilde{\rho}={}&\sum_{n\alpha\beta}\sum_{n'\alpha'\beta'}\gamma^{(-)}_{n\alpha\beta n'\alpha'\beta'}{\cal D}(\tilde{{\cal O}}_{n\alpha\beta},\tilde{{\cal O}}^{\dagger}_{n'\alpha'\beta'})\tilde{\rho}\nonumber\\
+{}&\sum_{n\alpha\beta}\sum_{n'\alpha'\beta'}\gamma^{(+)}_{n\alpha\beta n'\alpha'\beta'}\tilde{{\cal D}}(\tilde{{\cal O}}^{\dagger}_{n\alpha\beta},\tilde{{\cal O}}_{n'\alpha'\beta'})\tilde{\rho}\,,\label{liouvJC3}
\end{align}
where ${\cal D}(A,B)\tilde{\rho}=A\tilde{\rho}B-(BA\tilde{\rho}-\tilde{\rho}BA)/2$, and we have introduced the generalized transition rates
\begin{align}
\gamma^{(-)}_{n\alpha\beta n'\alpha'\beta'}={}&\gamma_m(\bar{n}_m+1)B_{n\alpha\beta}B_{n'\alpha'\beta'}\nonumber\\
{}&+\gamma_T(\bar{n}_T+1)\sigma_{n\alpha\beta}\sigma_{n'\alpha'\beta'},\\
\gamma^{(+)}_{n\alpha\beta n'\alpha'\beta'}={}&\gamma_m \bar{n}_m B_{n\alpha\beta}B_{n'\alpha'\beta'}\nonumber\\
{}&+\gamma_T \bar{n}_T\sigma_{n\alpha\beta}\sigma_{n'\alpha'\beta'}.
\end{align}

We employ standard projection operator techniques \cite{SIGardinerZoller,SIBreuerPetruccione}  to derive an effective master equation for  $\rho_s$, which we expand to second order in the interaction Liouvillian $\mathcal{L}_{\rm int}$ based on the hierarchy of timescales given in \eqref{eq:timescales}. The resulting effective dynamics takes place at a rate $\sim g^2/\kappa$ and is described by the interaction picture master equation
\begin{widetext}
\begin{align}
\dot{\tilde{\rho}}_{s}={}\tilde{\cal L}_{\rm JC}\tilde{\rho}_{s}&+g^2\sum_{n\alpha\beta}\sum_{n'\alpha'\beta'}B_{n\alpha\beta}B_{n'\alpha'\beta'}{\rm Re}\lbrace R(\omega_{n'\alpha'\beta'})\rbrace\left(\tilde{{\cal O}}_{n\alpha\beta}\tilde{\rho}_{s}\tilde{{\cal O}}^{\dagger}_{n'\alpha'\beta'}-\tilde{{\cal O}}^{\dagger}_{n\alpha\beta}\tilde{{\cal O}}_{n'\alpha'\beta'}\tilde{\rho}_{s}+{\rm h.c.}\right)\nonumber\\
{}&+g^2\sum_{n\alpha\beta}\sum_{n'\alpha'\beta'}B_{n\alpha\beta}B_{n'\alpha'\beta'}{\rm Re}\lbrace R(-\omega_{n'\alpha'\beta'})\rbrace\left(\tilde{{\cal O}}^{\dagger}_{n\alpha\beta}\tilde{\rho}_{s}\tilde{{\cal O}}_{n'\alpha'\beta'}-\tilde{{\cal O}}_{n\alpha\beta}\tilde{{\cal O}}^{\dagger}_{n'\alpha'\beta'}\tilde{\rho}_{s}+{\rm h.c.}\right)\nonumber\\
{}&+ig^2\sum_{n\alpha\beta}\sum_{n'\alpha'\beta'}B_{n\alpha\beta}B_{n'\alpha'\beta'}{\rm Im}\lbrace R(\omega_{n'\alpha'\beta'})\rbrace\left(\tilde{{\cal O}}_{n'\alpha'\beta'}\tilde{\rho}_{s}\tilde{{\cal O}}^{\dagger}_{n\alpha\beta}+\tilde{\rho}_{s}\tilde{{\cal O}}^{\dagger}_{n'\alpha'\beta'}\tilde{{\cal O}}_{n\alpha\beta}-{\rm h.c.}\right)\nonumber\\
{}&+ig^2\sum_{n\alpha\beta}\sum_{n'\alpha'\beta'}B_{n\alpha\beta}B_{n'\alpha'\beta'}{\rm Im}\lbrace R(-\omega_{n'\alpha'\beta'})\rbrace\left(\tilde{{\cal O}}^{\dagger}_{n'\alpha'\beta'}\tilde{\rho}_{s}\tilde{{\cal O}}_{n\alpha\beta}+\tilde{\rho}_{s}\tilde{{\cal O}}_{n'\alpha'\beta'}\tilde{{\cal O}}^{\dagger}_{n\alpha\beta}-{\rm h.c.}\right),\label{masterequation5}
\end{align}
\end{widetext}
where the response of the eliminated cavity enters via its spectrum 
\begin{align}
R(\omega)\equiv\frac{\kappa/2+i(\omega+\Delta_L)}{(\kappa/2)^2+(\omega+\Delta_L)^2}.
\end{align}

Complementing the above result, we also state the adiabatic relation between the cavity operator $a(t)$ and JC transition operators ${\cal O}_{n\alpha\beta}(t)$, which is useful for calculating the OM output spectrum below. Such a relation can be obtained by eliminating the cavity in the formalism of quantum Langevin equations \cite{SIGardinerZoller}, while again making use of the hierarchy of timescales given in \eqref{eq:timescales}. Omitting transients, we obtain the relation
\begin{align}
a(t)\approx{}&-\sum_{n\alpha\beta}\frac{igB_{n\alpha\beta}{\cal O}_{n\alpha\beta}(t)}{\kappa/2-i(\omega_{n\alpha\beta}+\Delta_L)}\nonumber\\
{}&-\sum_{n\alpha\beta}\frac{igB_{n\alpha\beta}{\cal O}^{\dagger}_{n\alpha\beta}(t)}{\kappa/2+i(\omega_{n\alpha\beta}-\Delta_L)}+{\rm noise}\label{adiabaticrel}\,,
\end{align}
where the noise contribution contains the vacuum input of the cavity that does not play a role in evaluating the spectrum below.

\subsection{Effective Jaynes-Cummings description with modified parameters}
\label{sec:effJC}

In addition to the conditions in \eqref{eq:timescales}, the setups of interest in this work can be assumed to satisfy
\begin{align}
\delta\omega\lesssim\lambda\ll\kappa.\label{extraconditionn}
\end{align}
To zeroth order in the small ratios $\delta\omega/\kappa$ and $\lambda/\kappa$, we can approximate $R(\pm\omega_{n\alpha\beta})\approx R(\pm\omega_m)$ in Eq.\,(\ref{masterequation5}) and thus pull these factors out of the sums. By subsequently reverting to the operators $b$ and $\sigma_-$, we see that the effective master equation assumes the same form as our original description of the dissipative JC model, i.e.,
\begin{align}
\dot{\rho}_{s}(t)={}&-\frac{i}{\hbar}[H_{\rm JC},\rho_{s}]{}+\bar{\gamma}(\bar{n}+1){\cal D}[b]\rho_{s}+\bar{\gamma}\bar{n}{\cal D}[b^{\dagger}]\rho_{s}\nonumber\\
{}&+\gamma_T(\bar{n}_T+1){\cal D}[\sigma_{-}]\rho_{s}+\gamma_T\bar{n}_T{\cal D}[\sigma_{+}]\rho_{s}.\label{masterequationJCeffective}
\end{align}
However, the mechanical parameters now incorporate the effect of the optical cavity: 
\begin{align}
\bar{\gamma}={}&\gamma_m+A^{(-)}-A^{(+)},\label{effective1}\\
\bar{n}={}&\frac{\bar{n}_m\gamma_m+A^{(+)}}{\gamma_m+A^{(-)}-A^{(+)}},\label{effective2}
\end{align}
where the optically induced cooling and heating rates are given as in Ref.\,\cite{SIWilsonRaeKippenberg2007} by
\begin{align}
A^{(\pm)}=\frac{g^2\kappa}{(\kappa/2)^2+(\Delta_L\mp\omega_m)^2}.
\end{align}
Note also that we neglected a small mechanical frequency shift of the order $\sim g^2/\omega_m\ll \omega_m$ in the JC Hamiltonian. 

Also under the extra condition (\ref{extraconditionn}), it is possible to simplify the adiabatic relation (\ref{adiabaticrel}) according to 
\begin{align}
a(t)\approx {}&\frac{-ig{}{}b(t)}{\kappa/2-i(\omega_m+\Delta_L)}+\frac{-ig{}{}b^{\dag}(t)}{\kappa/2+i(\omega_m-\Delta_L)}+{\rm noise}.\label{adiabaticsimple}
\end{align}

The effective master equation (\ref{masterequationJCeffective}), together with the adiabatic relation (\ref{adiabaticsimple}) are a convenient starting point for the analytical and numerical analysis of the system, since they are simple in form compared to Eqs.\,(\ref{masterequation5}) and (\ref{adiabaticrel}), and still describe all the physics we are interested in. Furthermore, they remain valid when adding a weak coherent microwave drive to the TLS, as described by $H_{{\rm TLS},\mu}=\hbar\Omega_{\mu}e^{i\omega_{\mu}t}\sigma_{-}+\hbar\Omega_{\mu}^{*}
e^{-i\omega_{\mu}t}\sigma_{+}$, with $|\Omega_{\mu}|\ll\lambda$. We therefore use these two equations to obtain the results on the phonon blockade effect displayed in Fig.\,3 of the main text, where we calculate the optical $g^{(2)}_c(0)\equiv\langle a^\dag a^\dag a a\rangle/\langle a^\dag a\rangle^2$ in the steady state of the master equation (\ref{masterequationJCeffective}). Here, a direct numerical treatment of the full problem is complicated by the fact that there is no interaction picture in which the full Liouvillian is time independent, such that \eqref{masterequationJCeffective} and \eqref{adiabaticsimple} are extremely helpful.

\subsection{Calculation of optomechanical output spectrum}

The output spectrum of the optical cavity is defined as  $S(\omega)\equiv (1/2\pi)\int_{-\infty}^{\infty}d\tau\ e^{-i\omega\tau}\langle a^{\dagger}_{\rm out}(\tau)a_{\rm out}(0)\rangle_{\rm ss}$ \cite{SIGardinerZoller}, where $a_{\rm out}(t)$ is the operator associated with the outgoing field, and the angular brackets $\langle ...\rangle_{\rm ss}$ denote the average in the steady state of the system. By using the input-output relation \cite{SIGardinerZoller}, the spectrum can be expressed in terms of the cavity mode operator:
\begin{align}
S(\omega)={}&\frac{\kappa}{\pi}\ {\rm Re}\left\lbrace \int_{0}^{\infty}d\tau e^{-i(\omega-\omega_L)\tau}\langle a^{\dagger}(\tau)a(0)\rangle_{\rm ss}\right\rbrace,\label{finalS}
\end{align}
where we have omitted a contribution $\sim\delta(\omega-\omega_L)$ due to the laser that drives the optomechanical system and causes the enhancement of the radiation pressure coupling.

By substituting the adiabatic relation (\ref{adiabaticrel}) into Eq.\,(\ref{finalS}), we see that we generally need two-time correlation functions between different transition operators, as for example $\langle {\cal O}^{\dag}_{n\alpha\beta}(\tau){\cal O}_{n'\alpha'\beta'}(0)\rangle_{\rm ss}$. Such correlation functions could in principle be calculated using \eqref{masterequation5}. However, note that for strong coupling and low enough temperatures the spectrum consists of non-overlapping peaks, each of which is associated with a transition between JC eigenstates. This situation is well described by the secular approximation \cite{SITianCarmichael92,SITian2011}, in which cross-terms between different transition operators are neglected based on the assumption $|\omega_{n\alpha\beta}-\omega_{n'\alpha'\beta'}|\gg|\gamma^{(-)}_{n\alpha\beta n'\alpha'\beta'}|$, for $(n\alpha\beta)\neq(n^\prime\alpha^\prime\beta^\prime)$. By means of this approximation we can greatly simplify \eqref{masterequation5} and obtain a result similar to the one stated in Ref.\,\cite{SITian2011}:
\begin{align}
\dot{\rho}_{s}{}&=-i[H_{\rm JC},\rho_{s}]\nonumber\\
{}&+\sum_{n\alpha\beta}\left\lbrace\Gamma^{(-)}_{n\alpha\beta}{\cal D}\left[{\cal O}_{n\alpha\beta}\right]\rho_{s}+\Gamma^{(+)}_{n\alpha\beta}{\cal D}[{\cal O}^{\dagger}_{n\alpha\beta}]\rho_{s}\right\rbrace.\label{masterequationSecular}
\end{align}
In the regime (\ref{extraconditionn}), the transition rates $\Gamma_{n\alpha\beta}^{(\pm)}$ in this master equation can be written as
\begin{align}
\Gamma^{(-)}_{n\alpha\beta}={}&\bar{\gamma} (\bar{n}+1)B_{n\alpha\beta}^2+\gamma_T(\bar{n}_T+1)
\sigma_{n\alpha\beta}^2,\label{simplerates1}\\
\Gamma^{(+)}_{n\alpha\beta}={}&\bar{\gamma}\bar{n}B_{n\alpha\beta}^2+\gamma_T \bar{n}_T\sigma_{n\alpha\beta}^2.\label{simplerates2}
\end{align}
Note that Eq.\,(\ref{masterequationSecular}) can also be obtained by applying the secular approximation directly to Eq.\,(\ref{masterequationJCeffective}).

The simplicity of \eqref{masterequationSecular} carries over to the equations of motion for the expectation values:
\begin{align}
\frac{d}{dt}\langle{\cal O}_{n\alpha\beta}(t)\rangle={}&-\left(\frac{\gamma_{n\alpha\beta}}{2}+i\omega_{n\alpha\beta}\right)\langle{\cal O}_{n\alpha\beta}(t)\rangle,\label{dynamicaleq}
\end{align}
where the effective decay rate for the transition with frequency $\omega_{n\alpha\beta}$ reads
\begin{align}
\gamma_{n\alpha\beta}={}&\sum_{\mu}\left\lbrace\Gamma^{(-)}_{n\alpha\mu}+\Gamma^{(-)}_{(n-1)\beta\mu}
+\Gamma^{(+)}_{(n+1)\mu\alpha}
+\Gamma^{(+)}_{n\mu\beta}\right\rbrace.\label{effectivedecayDef}
\end{align}
By means of the quantum regression theorem \cite{SIGardinerZoller,SIBreuerPetruccione}, this result enables us to calculate the two-time correlation functions needed to evaluate \eqref{finalS}.  As a result, the spectrum can be written as a sum over Lorentzians $l_{\omega}(\omega_0,\gamma_0)\equiv\gamma_0^2/(\gamma_0^2+(\omega-\omega_L-\omega_0)^2)$:
\begin{align}
S(\omega)={}&S_{\rm red}(\omega)+S_{\rm blue}(\omega)\\
         ={}&\sum_{n,\alpha,\beta}W^{\rm red}_{n\alpha\beta}\ l_{\omega}\left(-\omega_{n\alpha\beta},\gamma_{n\alpha\beta}/2\right)\nonumber\\
+{}&\sum_{n,\alpha,\beta}W^{\rm blue}_{n\alpha\beta}\ l_{\omega}\left(\omega_{n\alpha\beta},\gamma_{n\alpha\beta}/2\right),\label{spectrumquasi4}
\end{align}
where we have separated the contributions of the blue and red mechanical sidebands of the spectrum, introducing the weights:
\begin{align}
W^{\rm red}_{n\alpha\beta}={}&\frac{2A^{(+)}B^2_{n\alpha\beta}p_{(n-1)\beta}}{\pi\gamma_{n\alpha\beta}},\label{redweight}\\
W^{\rm blue}_{n\alpha\beta}={}&\frac{2A^{(-)}B^2_{n\alpha\beta}p_{n\alpha}}{\pi\gamma_{n\alpha\beta}}.\label{blueweight}
\end{align}
In Eqs.\,(\ref{redweight})-(\ref{blueweight}), $p_{n\alpha}\equiv\langle|n\alpha\rangle\langle n\alpha|\rangle_{\rm ss}$ denotes the steady state mean occupation of the JC eigenstate $|n\alpha\rangle$. The blue sideband term of Eq.\,(\ref{spectrumquasi4}) is the result quoted in Eq.\,(8) of the main text.

Using Eq.\,(\ref{masterequationSecular}) we find that the occupations $p_{n\alpha}$ satisfy a classical master equation, whose steady-state is determined by the recursion:
\begin{align}
{}&\sum_{\mu}\Gamma^{(-)}_{(n+1)\mu\alpha}p_{(n+1)\mu}-p_{n\alpha}\sum_{\mu}\left( \Gamma^{(-)}_{n\alpha\mu}+\Gamma^{(+)}_{(n+1)\mu\alpha}\right)\nonumber\\
{}&+\sum_{\mu}\Gamma^{(+)}_{n\alpha\mu}p_{(n-1)\mu}=0,\qquad n\ge 0,\label{recursion}
\end{align}
and where for compactness we defined $\Gamma^{(\pm)}_{1\alpha-}=0$, $\Gamma^{(\pm)}_{0\alpha\beta}=0$, $p_{0-}=0$ and $p_{(-1)\alpha}=0$. Since $\Gamma^{(\pm)}_{1++}=\Gamma^{(\pm)}_{1-+}$, it is easy to see that detailed balance solves Eq.\,(\ref{recursion}) for $n=0$:
\begin{align}
p_{1+}=p_{1-}=\left(\Gamma^{(+)}_{1++}/\Gamma^{(-)}_{1++}\right)p_{0+}.\label{recursion0}
\end{align}

Also noticing that the rates in Eqs.\,(\ref{simplerates1})-(\ref{simplerates2}) satisfy  $\Gamma^{(\pm)}_{n\alpha\beta}=\Gamma^{(\pm)}_{n\beta\alpha}$ and  that the sums $\sum_{\mu}\Gamma^{(\pm)}_{n\alpha\mu}$ are indepedent of $\alpha$, it is straightforward to show that Eq.\,(\ref{recursion}) implies
\begin{align}
p_{n-}=p_{n+},\qquad \forall n \ge 1.\label{PplusMinus}
\end{align}
As a result, Eq.\,(\ref{recursion}) simplifies and can be readily solved by detailed balance, yielding \cite{SITianCarmichael92}
\begin{align}
\frac{p_{(n+1)+}}{p_{n+}}=\left[\frac{\bar{\gamma}\bar{n}(2n+1)+\gamma_T \bar{n}_T}{\bar{\gamma}(\bar{n}+1)(2n+1)+\gamma_T(\bar{n}_T+1)}\right],\label{finalrecursion}
\end{align}
for $n\ge 0$. In order to fix the ground state occupation $p_{0+}$ one must impose the normalization condition $p_{0+}+2\sum_{n=1}^{\infty}p_{n+}=1$ and by this the complete steady state is obtained.

\subsection{Estimation of cross-over temperature}

Our final goal is to estimate the cross-over temperature $T_c$, at which the individual peaks in the optomechanical output spectrum $S_{\rm blue}(\omega)$ visible for $T\ll T_c$ merge into a single resonance at $\omega=\omega_{\rm blue}\equiv\omega_L+\omega_m$. Within the secular approximation and $\lambda\ll\kappa$, as in the previous subsection, the spectrum is always symmetric around $\omega_{\rm blue}$ and consists of many overlapping peaks as $T_c$ is aproached. In particular, only transitions \emph{within} the upper and lower branches of the JC ladder contribute, and we focus on the two most dominant peaks on either side of $\omega_{\rm blue}$, corresponding to the transitions $|n-\rangle\rightarrow|(n-1)-\rangle$ and $|n+\rangle\rightarrow|(n-1)+\rangle$, with maximum weights $W^{\rm blue}_{N++}=W^{\rm blue}_{N--}$. Here $N$ denotes the index $n$ for which the weights are maximal at given $T$. Then, the cross-over approximately occurs when these two highest Lorentzian peaks appreciably overlap, i.e. when the distance between their centers is approximately the same as the sum of their half widths:
\begin{align}
\omega_{N++}-\omega_{N--}\approx\left(\gamma_{N++}+\gamma_{N--}\right)/2.\label{criterium}
\end{align}
In the case of exact TLS-phonon resonance $\delta\omega=0$ and since $\omega_{N++}-\omega_{N--}=2\omega_{N++}$ 
and $\gamma_{N--}=\gamma_{N++}$, the cross-over condition (\ref{criterium}) can be simplified to
\begin{align}
2\lambda(\sqrt{N}-\sqrt{N-1})\approx \gamma_{N++}.\label{simplecriterium}
\end{align}

In order to find the index $N=N(T)$ as a function of temperature, one needs to maximize $W^{\rm blue}_{n++}$ in Eq.\,(\ref{blueweight}) as a function of $n$ for given $T$. Since a simple analytical expression for the ratio $p_{(n+1)+}/p_{n+}$ is given in Eq.\.(\ref{finalrecursion}), it is convenient to define $h_n\equiv W^{\rm blue}_{(n+1)++}/W^{\rm blue}_{n++}$, such that $N$ can be obtained from the condition $h_N\approx 1$, i.e., 
\begin{align}
h_{N}=\frac{p_{(N+1)+}}{p_{N+}}\frac{B^2_{(N+1)++}}{B^2_{N++}}\frac{\gamma_{N++}}{\gamma_{(N+1)++}}\approx 1.\label{Solvegn}
\end{align}
Therefore, by using Eqs.\,(\ref{matrixBs}), (\ref{effectivedecayDef}) and (\ref{finalrecursion}), one can in principle solve the above equation for $N$. However, before doing so, we can make further simplifications by restricting the parameters to the regime of experimental interest. Denoting by $\bar{n}_m^{\rm c}$ the mechanical Bose occupation of the bath at the cross-over temperature $T_c$, we write the conditions:
\begin{align}
A^{(+)}\ll\gamma_m \bar{n}_m^{\rm c}\ll A^{(-)}, \label{further1}\\
A^{(-)}\lesssim\gamma_T,\label{further12}\\
\bar{n}_m^{\rm c}\gg 1\,.\label{further2}
\end{align}
Here, \eqref{further1} are the conditions for optomechanical ground state cooling, while \eqref{further12} states that the TLS decay rate is the largest decoherence rate. In addition, we expect that for strong coupling $\lambda\gg\gamma_T$, the cross-over happens at temperatures corresponding to large mean occupations of the bath, which is expressed by \eqref{further2}. The above conditions imply that $\bar{\gamma}\approx A^{(-)}\lesssim\gamma_T$ and $\bar{n}\approx(\gamma_m \bar{n}_m^{\rm c})/A^{(-)}\ll 1\ll \bar{n}_m^{\rm c}$. As a consequence, we are able to perform a Taylor expansion of Eq.\,(\ref{Solvegn}) in powers of $(NA^{(-)})/(\gamma_T\bar{n}_m^{\rm c})\ll 1$, which to zeroth order gives a quadratic equation for $N$:
\begin{align}
c(c-1)^2N^2-\left(\frac{c+1}{2}\right)^4+{\cal O}\left(\frac{NA^{(-)}}{\gamma_T\bar{n}_m^{\rm c}}\right)=0.\label{simplifiedEQ}
\end{align}
Here $c$ is a dimensionless constant of the order one, given by
\begin{align}
c={}&1+\frac{1}{\bar{n}_m^{\rm c}}+\frac{2A^{(-)}}{\bar{n}_m^{\rm c}\gamma_T}+{\cal O}\left(\left[\frac{A^{(-)}}{\gamma_T\bar{n}_m^{\rm c}}\right]^2,\left[\frac{1}{\bar{n}_m^{\rm c}}\right]^2\right).\label{cexpression}
\end{align}
With the help of \eqref{cexpression} we can easily solve \eqref{simplifiedEQ} for $N$, yielding
\begin{align}
N={}&\frac{\bar{n}_m^{\rm c}}{1+2A^{(-)}/\gamma_T}+\frac{1}{2}+{\cal O}\left(\frac{A^{(-)}}{\gamma_T\bar{n}_m^{\rm c}},\frac{1}{\bar{n}_m^{\rm c}}\right)\\
\approx{}&\frac{\bar{n}_m^{\rm c}}{1+2A^{(-)}/\gamma_T}\gg 1.\label{simpleNNN}
\end{align}

Finally, we can plug our estimate (\ref{simpleNNN}) in the cross-over condition (\ref{simplecriterium}) and solve for $\bar{n}_m^{\rm c}$. By applying the approximations (\ref{further1})-(\ref{further2}) to $\gamma_{N++}$ given in (\ref{effectivedecayDef}) and assuming from Eq.\,(\ref{simpleNNN}) that $N\gg 1$, we are left with the equation,
\begin{align}
\gamma_T \bar{n}_m^{\rm c}\left(1+\frac{A^{(-)}/\gamma_T}{1+2A^{(-)}/\gamma_T}\right)\approx{}&\frac{\lambda}{2\sqrt{N}}.\label{almost}
\end{align}
If we now square Eq.\,(\ref{almost}) and use expression (\ref{simpleNNN}), we obtain the following analytical estimate for $\bar{n}_m^{\rm c}$:

\begin{align}
\bar{n}_m^{\rm c}\approx\left(\frac{\lambda}{2\gamma_T}\right)^{2/3}\frac{\left(1+2A^{(-)}/\gamma_T\right)\hspace{4mm}}{\left(1+3A^{(-)}/\gamma_T\right)^{2/3}}.\label{finalnmc}
\end{align}
This result consistenly satisfies our initial assumption, $\bar{n}_m^{\rm c}\gg 1$, provided the strong coupling condition, $\lambda\gg\gamma_T$, is fulfilled. By approximating the Bose distribution for large occupations $\bar{n}_m^{\rm c}\gg 1$, the corresponding cross-over temperature $T_c$ reads
\begin{equation}
T_c\approx \frac{\hbar\omega_m}{k_B}\bar n_m^c\approx \frac{\hbar\omega_m}{k_B}\left(\frac{\lambda}{2\gamma_T}\right)^{2/3}\frac{\left(1+2A^{(-)}/\gamma_T\right)\hspace{4mm}}{\left(1+3A^{(-)}/\gamma_T\right)^{2/3}},\label{SIanalyticalcross}
\end{equation}
which is the result quoted in Eq.\,(9) of the main text.

\setcounter{enumiv}{0}

\end{document}